\documentclass[11pt,twoside]{article}
\usepackage{asp2010}

\resetcounters

\begin{document}
\title{How important are metal-poor AGB stars as cosmic dust producers?}
\author{L. Mattsson$^1$, B. Aringer$^2$ and A. C. Andersen$^3$
\affil{$^1$Nordita, KTH Royal Inst. of Technology and Stockholm University, Sweden}
\affil{$^2$Department of Astronomy, University of Vienna, Austria}
\affil{$^3$Dark Cosmology Centre, NBI, University of Copenhagen, Denmark}
}

\begin{abstract}
The efficiency of dust formation in oxygen-rich AGB stars should (in theory) be metallicity dependent since they are not producing their own raw material for dust production. Metal-poor carbon stars may not be very efficient dust producers either, because of more radiative heating of the grains forming in their atmospheres. We have just confirmed that inefficient dust and wind formation in simulations of metal-poor carbon stars is a real physical effect, albeit within the limitations of our simulations. Taken at face value, this implies that the amount of dust supplied by low-metallicity AGB stars to the build up of the cosmic dust component is clearly limited. Consequently, one may also ask how large a contribution AGB stars can make in general, when compared to recent observations of cosmic dust, which are suggesting major contributions from other sources?
\end{abstract}

\section{Introduction}
There is plenty of evidence for the existence of very dusty galaxies at high redshifts ($z>4$). Several objects show dust-to-gas ratios which are as high, or possibly higher, than in local galaxies \citep[see, e.g.,][and references therein]{Mattsson11}. At such early times, the number of AGB stars is still small and not even with the most generous dust yield estimates, the AGB stars can make a contribution that compares to the observed dust masses \citep{Michalowski10}. The only reasonable way to explain the dusty galaxies in the early Universe is therefore either dust production in the short-lived high-mass stars (which indeed produces most of the metals) or rapid dust growth in the interstellar medium (ISM).

Recent detections of large masses of cold dust \citep[e.g.][]{Matsuura11,Gomez12} in supernova remnants suggest AGB stars may, relatively speaking, be less important than previously anticipated. Carbon stars (C stars) may have produced most of the carbon that we have today \citep{Carigi05,Mattsson10b}, but not necessarily the carbonaceous dust. Here we will discuss the role of AGB stars in general and C stars in particular. It will be argued that, besides time-scale issues, really metal poor C stars may form dust rather inefficiently despite they produce their own raw material for dust formation.

\section{The general problem with AGB stars as major dust producers}
A supernova (SN) can produce around $0.5\,M_\odot$ of dust \citep[see, e.g.,][]{Matsuura11,Gomez12}. We can assume that a typical AGB star ejects about $1\,M_\odot$ of gas, in which the mass fraction of condensable material is $10^{-4}\dots 10^{-3}$. From modelling \citep[e.g.][]{Mattsson08,Mattsson10,Ventura12,Nanni13} it seems clear that the mean degree of dust condensation is $\langle f_{\rm c} \rangle < 0.5$ (often $<0.1$, see Fig.~1), which can be understood as a consequence of the fact that as soon as a sufficient amount of dust is present to sustain a dust-driven outflow, the grains will move out of the dust-formation zone and stop growing. Combining the above facts, it is obvious that one needs a few thousand AGB stars to produce as much dust as does one supernova. The number of AGB stars is of course much greater than the number of SNe, but the AGB/SN ratio expected for a Galactic IMF is sooner $\sim 50$ (or less, depending on the mass cuts), than $\sim 5000$, which leaves us with a gap of about two orders of magnitude. 

Another potential problem arises from the fact that the AGB stars supplying dust to the ISM may have formed at really low metallicity, especially in young stellar populations. The silicate-dust production of oxygen-rich AGB stars (a.k.a. M giants), must have an upper limit due to metallicity because these stars are not producing their own raw material for dust production, which is indeed also seen in dust-formation models \citep{Ventura12,Nanni13}. Thus, carbonaceous dust from C stars should dominate the injection of dust into the ISM at low metallicity, if it is not SNe or dust condensation in the ISM. But then there are time-scale issues. Obviously, C stars cannot contribute dust at very high redshifts ($z>4$) since the Universe was too young for a majority of them to have evolved, but also in low-metallicity environments there is a similar problem: the dominant stellar population is young - typically as young as those at high $z$ - which should result in a lower frequency of C stars.

Let us elaborate a bit on time-scale issues by considering the following.  In the Small Magellanic Cloud (SMC), the total mass of dust in the ISM is a few times $10^5\,M_\odot$ \citep[$3\times 10^5\,M_\odot$ according to][]{Leroy07}. If one assumes $25-50$\% (by mass) of carbonaceous dust, one could simply say there is $\sim 10^5\,M_\odot$ of such dust present in the SMC. According to \citet{Matsuura13} the dust-injection rate by C stars in the SMC is $4\times 10^{-6}\,M_\odot$~yr$^{-1}$ and \citet{Boyer12} reports a rate of $7.5\dots 8.4\times 10^{-7}\,M_\odot$~yr$^{-1}$. Taken at face value these rates then imply dust-injection time-scales which are $25$~Gyr and $119\dots 133$~Gyr, respectively. That is, if the carbonaceous dust in the SMC is due to C stars only, this crude estimate suggests it would take longer than the age of the Universe to produce that amount. To be reasonable hypothesis, this time-scale estimate must be {\it significantly less} than the age of the Universe, which seems to require a higher dust-injection rate and/or lower total dust mass in the SMC.

\section{New results on dust and wind formation at low metallicity}
We have computed a small grid of models (Mattsson et al., in prep.) which all correspond to a star of mass $M_\star = 1\,M_\odot$ and luminosity $\log(L_\star/L_\odot)=4.0$, but different metallicities and effective temperatures (see Fig.~1). All models also have the same absolute abundances of free carbon, which excludes the possibility that the increased wind speed is due to the accessibility of raw material for dust formation. 

The main result is a somewhat unexpected trend with metallicity: sustaining dust-driven outflows from C-stars becomes increasingly more difficult with decreasing metallicity until a critical metallicity is reached where the mean degree of dust condensation $f_{\rm c}$ drops (see Fig.~1), despite the available carbon to form dust from. Thus, oxygen-rich AGB stars may not produce dust-driven winds due to their lack of raw material for dust formation; C-stars may not produce winds due to atmospheric conditions that lead to too little dust production.
Scaling the atmospheric abundances down to $1/100\,Z_\odot$ no models produce a sustained outflow, except for an unclear case shown in Fig.~1 as an upper limit, where the wind slows down and finally collapses. This could be an initial instability that leads to sufficient dust formation, but which fades away as the system is dissipating. We would like to caution the reader this is likely not the beginning of an episodic outflow pattern, but since the time step drops to almost zero when the collapse occurs, the time series could not be continued.

The lower absolute abundances of elements other than carbon affect molecular opacities and, consequently, the physical conditions in the atmosphere. Models with enhanced abundances of oxygen and nitrogen appear to form dust and winds more easily (see Fig.~1, right panel), although not quite as in the case of a solar composition. For two of the three successful models we must regard the mass-loss rates we obtain as upper limits because the outflows are episodic and one cannot be sure these, possibly semi-chaotic, cycles are stable.

\begin{figure}
  \resizebox{\hsize}{!}{
\includegraphics{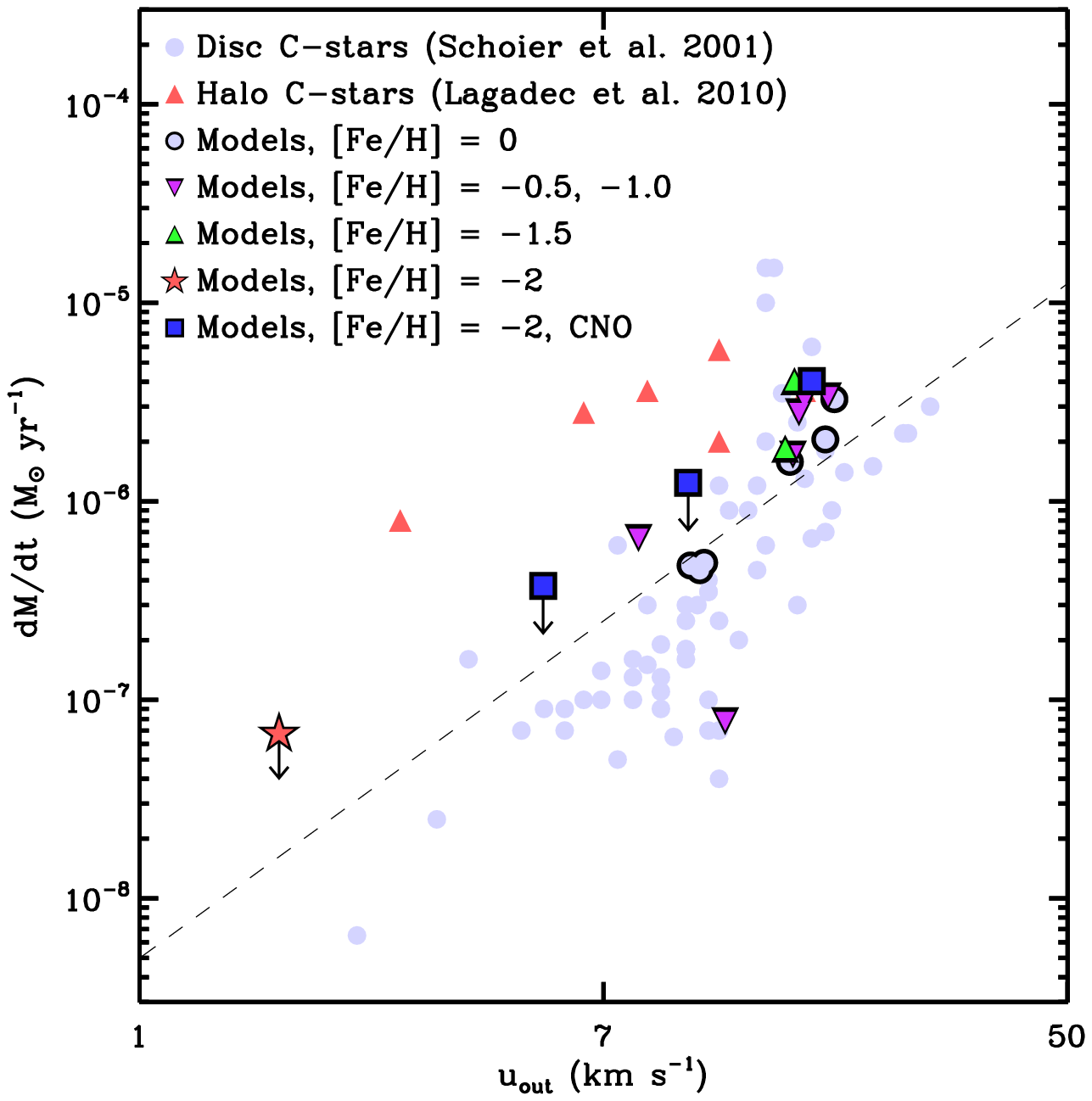}
\includegraphics{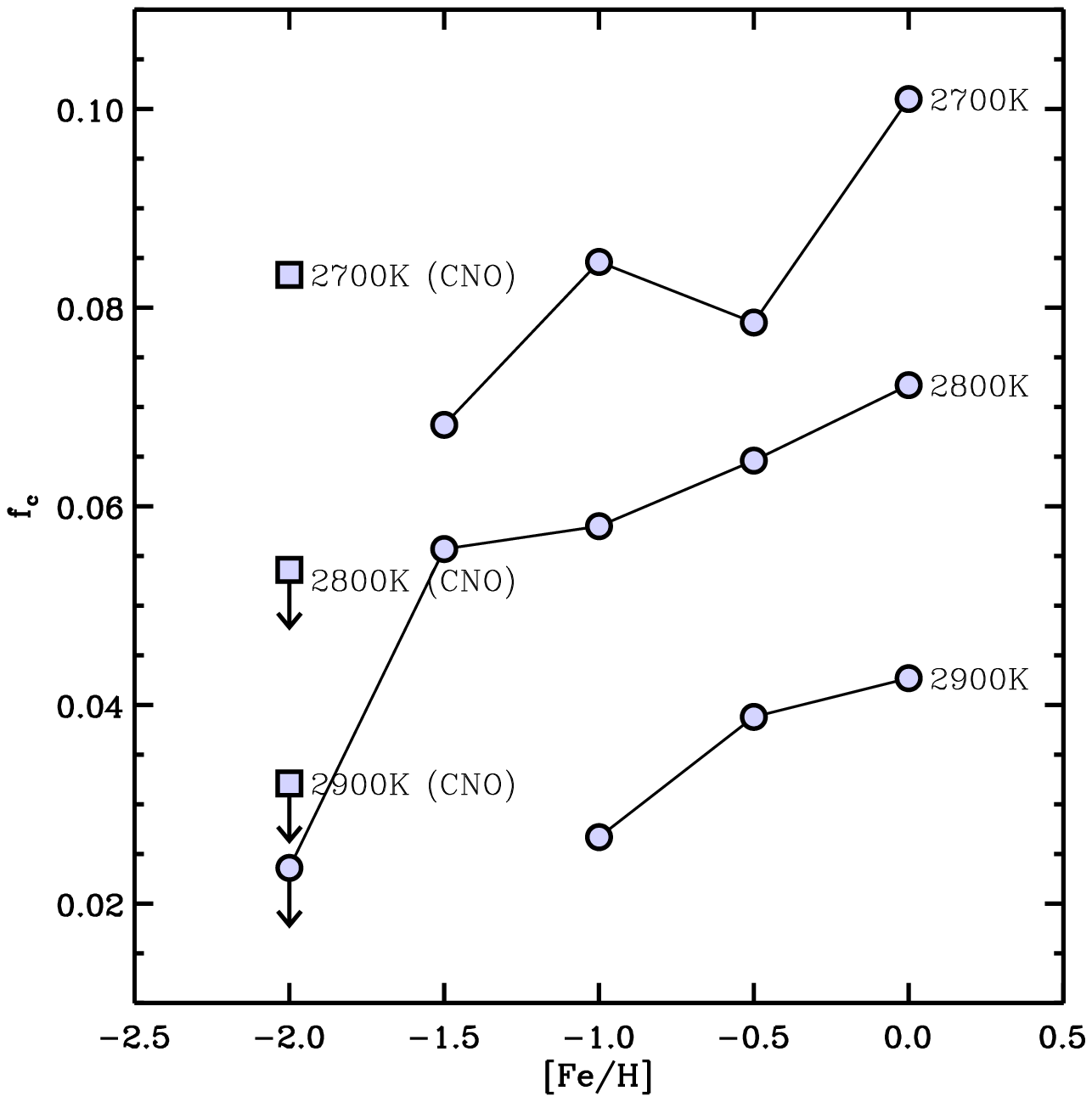}
}
\caption{\label{lowz} Left: mass-loss rate as a function of wind speed for models and observed C stars. Right: mean degree of dust condensation as a function of metallicity.
}
\end{figure}

At the lowest metallicities, the $P-T$ structures deviate from the more metal-rich models just above the photosphere, in a region where many molecules, and C$_2$H$_2$ in particular, usually form. A similar deviation is seen in the deeper layers of the atmospheres. Since the pressure is lower and the temperature somewhat higher, less of the important cooling agent and growth species C$_2$H$_2$ is formed. The same is true for many diatomic molecules too (especially C$_2$) which means that C$_2$ is not taking over as the source of opacity as one might have expected. Perhaps even more important is that molecules containing nitrogen, such as CN and HCN, will be scarce simply because there is very little nitrogen when the overall metallicity is scaled down below [Fe/H]~$ = -1$. At shorter wavelengths, there will also be little absorption due to atomic species. Hence, there is an `opacity deficiency' in the metal-poor models.

With these opacities reduced, newly formed dust grains will be exposed to more radiation and thus heat up to temperatures high enough for many grains to sublimate, or at least grow inefficiently. The sublimation rate is roughly $R_{\rm sub} \approx 0.015\,P_{\rm C}(T_{\rm d})\, T_{\rm d}^{-1/2}$, where the vapour saturation pressure for atomic carbon $P_{\rm C}$ can be estimated by the formula $\ln(P_{\rm C}) = 32.89 - 86300/T_{\rm d}$ \citep{Gail88}. Clearly, $R_{\rm sub}$ has a steep dependence on the dust temperature $T_{\rm d}$. More precisely, increasing the dust temperature by a few percent can raise the sublimation rate a couple of orders of magnitude. Thus, even a moderate increase in the efficiency of radiative heating of the dust can lead to a sublimation rate that supersedes the condensation rate, resulting in a low (or even zero) mean degree of dust condensation. The low abundance of molecules including nitrogen seems to play key role here. Low metallicity ([Fe/H]$ = -2$) models with elevated nitrogen and oxygen abundances (marked CNO in Fig.~1) show dust formation, while test runs with solar composition, except that [N/Fe]$ = -2$, leads to essentially no dust formation at all. At the time of writing we are not sure how to interpret this result.

There are also reasons to believe that metal-poor C-stars are hotter than their more metal-rich counterparts \citep{Sloan09}. If this is generally the case, the chance of having efficient dust condensation is small, from a theoretical perspective, as the overall gas/dust temperature will decrease with the effective temperature. Observations of dust around metal-poor C-stars in the halo \citep[e.g.][]{Sloan09,Lagadec10} may thus seem inconsistent with our model. C stars do indeed form significant amounts of dust also at [Fe/H]$\sim -2$ (see Boyer et al. in this volume), but the total {\it dust yield} may still be less than at solar metallicty, because the average degree of dust condensation, taken over the whole AGB, would be lower.

\section{AGB evolution and mass loss}
Models of AGB evolution which include dust formation rely on simple prescriptions for the mass-loss rate as a function of the basic stellar parameters \citep{Ventura12,Nanni13}. Imposing a certain mass-loss rate and requiring that the outflow should be dust-driven, will essentially make the efficiency of dust formation an input parameter and not a result of the modelling. Needless to say, this must affect the dust yield. So how would the dust yield be affected by a more consistent treatment of the mass loss? The dust yield is an integral quantity, where $f_{\rm c}$ and the fraction of free condensable material $\tilde{X}_{\rm c}$ are convolved with the mass-loss rate $\dot{M}$,
\begin{equation}
Y_{\rm d} = \int_{\tau_{\rm AGB}} f_{\rm c}(t)\, \tilde{X}_{\rm c}(t)\,\dot{M}(t)\,dt,
\end{equation}
where $\dot{M}(t) = f(f_{\rm c}, \tilde{X}_{\rm c}, \dots)$, which generates feedback and nonlinearity. The integral, $Y_{\rm d}$, can vary a lot depending on how the quantities given above evolve. On top of that, the average {\it grain size} is sometimes decisive for whether a dust-driven outflow develops or not in some cases \citep{Mattsson11b}. In short: stellar evolution models need a physically consistent mass-loss prescription. Estimating dust yields requires that dust formation efficiency leads to a certain mass-loss rate and not the other way around. In particular, as our models of C stars suggest, dust may form without leading to dust-driven outflows (not quite enough dust to reach critical acceleration), which means the amount of dust being expelled by the star can be insignificant despite ongoing dust formation.

\section{Conclusions and open questions}
In the big picture, AGB stars appear to be relatively minor dust producers. The quantitative uncertainties regarding, e.g., the efficiency of dust condensation in SN remnants are still considerable, but detections of {\it cold} dust have shown unambiguously that SNe can produce significant amounts dust. To compete with SNe, AGB stars must produce much more dust than they are likely able to.

Oxygen-rich AGB stars produce mainly silicate dust, but the production is limited by the initial metallicity.  C stars, on the other hand, produce their own raw material for dust production (carbon). However, our recent simulations show that metal poor C stars may form dust less efficienctly. The reason is that at really low metallicity ([Fe/H]$\sim -2$), the molecular gas opacity is reduced to a minimum and the grains are not shielded from radiative heating as much as at higher metallicity, which in turn increases the sublimation rate at the expense of net condensation. 

The implications of the above are interesting and raises several questions: \\[-6mm] 
\begin{itemize}
\item If AGB stars provide only a fraction of the cosmic dust, why do presolar grains show isotopic signs of being of mostly AGB origin \citep[see, e.g.,][]{Nittler03}?\\[-6mm] 
\item If the time-scale issue exclude C stars as significant  dust producers in low-metallicity environments, where interstellar grain growth is also inefficient \citep{Asano13,Mattsson12}, could one then argue that dust in really metal-poor stellar populations is essentially SN dust?\\[-6mm] 
\item Really metal poor C stars may also undergo relatively long phases without any strong dust-driven outflows, which should have significant effects on the late-stage evolution of low these stars. It is unclear if dust-driven winds can actually remove the whole envelope within the required time in order to form a white dwarf, so what is the fate of these stars?\\[-6mm]  
\end{itemize}


\end{document}